\documentclass[11pt]{article}
\usepackage{amssymb}
\usepackage{graphicx,color}
\usepackage{amsmath}
\usepackage{amsbsy}
\usepackage{epsfig}
\newtheorem{lemma}{Lemma}[section]

\newtheorem{theorem}{Theorem}[section]

\newtheorem{remark}{Remark}[section]

\def\proclaim#1{\par \bigskip\noindent {\bf #1}\bgroup\it\ }
\def\endproclaim{\egroup\par\bigskip}
\def\proof{\par\noindent{\bf Proof.} \;}

\newbox\TempBox \newbox\TempBoxA
\newcommand{\non}{\nonumber \\}

\def\text#1{\mbox{\rm #1}}

\def\underwiggle 1{
\ifmmode\setbox\TempBox=\hbox{$ 1$}\else\setbox\TempBox=\hbox{ 1}\fi
\setbox\TempBoxA=\hbox to \wd\TempBox{\hss\char'176\hss}
\rlap{\copy\TempBox}\smash{\lower9pt\hbox{\copy\TempBoxA}} }

\newcommand{\be}{\begin{eqnarray}}
\newcommand{\ee}{\end{eqnarray}}
\newcommand{\by}{\begin{eqnarray*}}
\newcommand{\ey}{\end{eqnarray*}}
\newcommand{\bn}{\begin{enumerate}}
\newcommand{\en}{\end{enumerate}}
\newcommand{\bi}{\begin{itemize}}
\newcommand{\ei}{\end{itemize}}
\newcommand{\bds}{\begin{description}}
\newcommand{\eds}{\end{description}}
\newcommand{\bcen}{\begin{center}}
\newcommand{\ecen}{\end{center}}

\def\dconv{\stackrel{\mbox{d}}{\longrightarrow}}

\def\pconv{\stackrel{\mbox{P}}{\longrightarrow}}

\begin{document}

\title{\bf Asymptotic Inference for AR(1) Penal Data\footnote{This research was partly supported by the Department of Education of Zhejiang Province (N20140202).
\newline
\hspace*{5mm}Address correspondence to Tianxiao Pang, School of Mathematical Sciences, Yuquan Campus, Zhejiang University, Hangzhou 310027, P.R.China; email: txpang@zju.edu.cn.
}\\}
\date{}

\maketitle
\begin{center}
\vskip -1cm {{\small\sc Jianfei Shen and Tianxiao Pang}}
{\small\centerline{Zhejiang University}}
 \end{center}
\maketitle

\bigskip
{\bf Abstract.} A general asymptotic theory is given for the panel data AR(1) model with time series independent in different cross sections. The theory covers the cases of stationary process, nearly non-stationary process, unit root process, mildly integrated, mildly explosive and explosive processes. It is assumed that the cross-sectional dimension and time-series dimension are respectively $N$ and $T$. The results in this paper illustrate that whichever the process is, with an appropriate regularization, the least squares estimator of the autoregressive coefficient converges to a normal distribution with rate at least $O(N^{-1/3})$. Since the variance is the key to characterize the normal distribution, it is important to discuss the variance of the least squares estimator. We will show that when the autoregressive coefficient $\rho$ satisfies $|\rho|<1$, the variance declines at the rate $O((NT)^{-1/2})$, while the rate changes to $O(N^{-1/2}T^{-1})$ when $\rho=1$ and $O(N^{-1/2}\rho^{-T+2})$ when $|\rho|>1$. $\rho=1$ is the critical point where the convergence rate changes radically. The transition process is studied by assuming $\rho$ depending on $T$ and going to $1$. An interesting phenomenon discovered in this paper is that, in the explosive case, the least squares estimator of the autoregressive coefficient has a standard normal limiting distribution in panel data case while it may not has a limiting distribution in univariate time series case.\\



\newpage

\section{Introduction}
\setcounter{equation}{0}

Dynamic models are useful in modeling time series data and have been well studied in the past few decades. One of the dynamic models is the AR(1) model which is given by
\begin{eqnarray}\label{ar1}
y_t=\rho y_{t-1}+\varepsilon_{t},\qquad  t=1,2,...,T.
\end{eqnarray}
For simplicity, in the sequel, we assume $y_0=0$ and $\{\varepsilon_t, t\ge 1\}$ are independent and identically distributed (i.i.d.) random variables with $E[\varepsilon_1]=0$ and $E[\varepsilon_1^2]=1$.

Although the model (\ref{ar1}) is simple, it is very useful and important in time series and econometrics literatures since the model can be used to model some kinds of stationary or non-stationary time series data. The parameter $\rho$ is the main concern in the model (\ref{ar1}) since whether the model is stationary is determined by the value of $\rho$. It is well-known that the necessary and sufficient condition for the stationarity of (\ref{ar1}) is $|\rho|<1$ if $y_0$ is an appropriate random variable or a constant. In this paper, we still call (\ref{ar1}) with $y_0=0$ a stationary AR(1) model when $|\rho|<1$, since this modification will not change the limiting distribution of the least squares estimator (LSE) of $\rho$ which is given by
\begin{align}\label{lse}
\hat{\rho}=\frac{\sum_{t=1}^T y_{t}y_{t-1}}{\sum_{t=1}^T y_{t-1}^2}.
\end{align}
For the stationary AR(1) model, Mann and Wald (1943) proved that
\begin{align*}
\frac{\sqrt{T}}{\sqrt{1-\rho^2}}(\hat{\rho}-\rho)\dconv N(0,1).
\end{align*}
When $\rho$ stisfies $|\rho|>1$, model (\ref{ar1}) is non-stationary and is called the explosive AR(1) model. For this model, Anderson (1959) showed that if $\varepsilon_t$'s are independent and normal distributed random variables, then
\begin{align*}
\frac{\rho^T}{\rho^2-1}(\hat{\rho}-\rho)\dconv C,
\end{align*}
where $C$ is a standard Cauchy variate. However, for general $\varepsilon_t$'s, Anderson (1959) showed that the limiting distribution of $\hat{\rho}$ may not exist. The interesting case is $\rho=1$, the corresponding AR(1) model is called the unit root model in econometrics. For this model, the central limit theorem is no longer applicable when exploring the limiting distribution of $\hat{\rho}$. Instead, by applying functional central limit theorem, White (1958) and Rao (1978) showed that
\begin{align*}
T(\hat{\rho}-\rho) \dconv \frac{\frac{1}{2}\left[W^2(1)-1\right]}{\int_0^1 W^2(t)dt},
\end{align*}
where $\{W(t), 0\leq t\leq 1\}$ is a standard Wiener process. The limiting distribution is not standard. Noting that $P(\hat{\rho}-\rho \leq 0)=P(W^2(1)\leq 1)\approx 0.684$, the limiting distribution is not even symmetric.

In order to bridge the gaps of asymptotic theories between the stationary AR(1) model and the unit root model, Chan and Wei (1987) and Phillips (1987) independently studied the following model called nearly non-stationary AR(1) model:
\begin{eqnarray}\label{cw}
y_t=\rho y_{t-1}+\varepsilon_{t}, \qquad y_0=0, \qquad  \rho=\rho_T=1-c/T,\qquad t=1,2,...,T,
\end{eqnarray}
where $c$ is a fixed constant. Of late, in order to bridge the gaps of asymptotic theories between the unit root model and the explosive AR(1) model, Phillips and Magdalinos (2007) studied the following AR(1) model:
\begin{eqnarray}\label{pm}
y_t=\rho y_{t-1}+\varepsilon_{t},\qquad y_0=0, \qquad \rho=\rho_T=1-c/k_T,\qquad t=1,2,...,T,
\end{eqnarray}
where $c$ is a fixed constant and $k_T$ is a sequence of positive constants increasing to $\infty$ such that $k_T=o(T)$. Model (\ref{pm}) with $c>0$ and with $c<0$ is called mildly integrated AR(1) model and mildly explosive AR(1) model respectively according to Phillips and Magdalinos (2007).

In models (\ref{cw}) and (\ref{pm}), we denote $\hat{\rho}_T$ the LSE of $\rho_T$ and also suppose that $\{\varepsilon_t, t\ge 1\}$ are i.i.d. random variables with $E[\varepsilon_1]=0$ and $E[\varepsilon_1^2]=1$. It is worth noting that the limiting distributions of $\hat{\rho}_T$ are different from those in the stationary AR(1) model, unit root model and explosive model. Specifically, Chan and Wei (1987) proved that when $\rho=\rho_T=1-c/T$ with $c\in R$,
\begin{eqnarray*}
T(\hat{\rho}_T-\rho_T)\dconv \frac{2c}{b}\frac{\int_0^1 (1+bt)^{-1}W(t)dW(t)}{\int_0^1 (1+bt)^{-2}W^2(t)dt},
\end{eqnarray*}
where $b=e^{2c}-1$ ($\frac{2c}{b}$ in the above limiting distribution is replaced by $1$ if $c=0$), while Phillips and Magdalinos (2007) proved that when $\rho=\rho_T=1-c/k_T$ with $c>0$,
\begin{eqnarray*}
\sqrt{Tk_T}(\hat{\rho}_T-\rho_T)\dconv N(0, 2c)
\end{eqnarray*}
and when $\rho=\rho_T=1-c/k_T$ with $c<0$,
\begin{eqnarray*}
[k_T\rho_T^T/(-2c)](\hat{\rho}_T-\rho_T)\dconv C,
\end{eqnarray*}
where, as before, $C$ stands for a standard Cauchy variate.

It is clear that the limiting distribution of the LSE of $\rho$ varies in AR(1) models under different assumptions on $\rho$. Further, one can find that the limiting distribution is not standard in nearly non-stationary AR(1) model which includes the unit root model as a special case. This is harmful for making further statistical inferences, for example, confidence intervals of $\rho$.

However, with the penal data, the results may be extremely simple. A panel data set is the one that follows a given sample of individuals over time, and thus provides multiple observations on each individual in the sample. A panel data AR(1) model is formulated by
\begin{align}\label{1.5}
y_{it}=\rho y_{i,t-1}+\varepsilon_{it},\qquad  t=1,2,...,T, \qquad i=1,2,...,N,
\end{align}
where, for simplicity, we suppose in this paper that $y_{i0}=0$ for any $i\ge 1$ and $\{\varepsilon_{it}, i\ge 1, t\ge 1\}$ are i.i.d. random variables with $E[\varepsilon_{11}]=0$ and $E[\varepsilon_{11}^2]=1$. The dimension of individual, $N$, is usually called cross-sectional dimension. There is no common effect on individuals in the model (\ref{1.5}). Thus each individual generates an independent time series and the central limit theorem may be applied on cross-sectional dimension. For this panel data AR(1) model, Levin and Lin (1992) proved that, when $\rho=1$ (unit root case) and an additional moment condition is fulfilled, that is, $E|\varepsilon_{11}|^{2+\lambda}<\infty$ for some $\lambda>0$, one has
\begin{eqnarray}\label{1.6}
\sqrt{N}T(\hat{\rho}-\rho) \dconv  N(0,2),  \qquad N, T\rightarrow \infty.
\end{eqnarray}
Obviously, the limiting distribution of $\hat{\rho}$ in panel data unit root model is simpler than that in univariate time series unit root model. What is more important is the former is standard while the latter is not. This comparison motivates us to study other panel data AR(1) models.

Therefore, the aim of this paper is to study the limiting distribution of the LSE of $\rho$ in various panel data AR(1) models. We are interested in the following question: whether, like the panel data unit root case, all the limiting distributions are normal in panel data stationary, nearly non-stationary, mildly integrated, mildly explosive and explosive cases.

The rest of the paper is organized as follows. We will extend the conclusion (\ref{1.6}) to general cases for $\rho\in R$ in Section 2, and provide some applications in Section 3. Note that, in Section 3, all the limiting distributions have the form of normal distribution only with different rates of convergence. When $\rho=1$, our result coincides with that in Levin and Lin (1992), but the moment condition $E[|\varepsilon_{11}|^{2+\lambda}]<\infty$ for some $\lambda>0$ is replaced by a more weaker one, that is, $E[\varepsilon_{11}^2]<\infty$, in our paper.\\

\section{Asymptotics for the LSE of $\rho$}
\setcounter{equation}{0}

Consider the panel data AR(1) model:
\begin{align}\label{pdmodel}
y_{it}=\rho y_{i,t-1}+\varepsilon_{it},\qquad  t=1,2,...,T, \qquad i=1,2,...,N,
\end{align}
where $y_{i0}=0$ for all $i\ge 1$ and the innovations $\{\varepsilon_{it}, i\ge 1, t\ge 0\}$ are i.i.d. random variables with $E[\varepsilon_{11}]=0$ and $E[\varepsilon_{11}^2]=1$. In this model, the LSE of $\rho$ is
\begin{eqnarray}\label{pdlse}
\hat{\rho}=\frac{\sum_{i=1}^N\sum_{t=1}^T y_{it}y_{i,t-1}}{\sum_{i=1}^N\sum_{t=1}^T y_{i,t-1}^2}.
\end{eqnarray}
It is true that
\begin{align}\label{rhohat}
\hat{\rho}-\rho=\frac{\sum_{i=1}^N \sum_{t=1}^T y_{i,t-1}\varepsilon_{it}}{\sum_{i=1}^N\sum_{t=1}^T y_{i,t-1}^2}.
\end{align}
To obtain a non-degenerated limiting distribution for (\ref{rhohat}), we can apply the central limit theorem to the numerator and the law of the large numbers to the denominator, respectively. Before doing so, we need to put the normalizing constants on $\sum_{t=1}^T y_{i,t-1}\varepsilon_{it}$'s and $\sum_{t=1}^T y_{i,t-1}^2$'s such that they become bounded in probability. The following is our main result in this section.


{\theorem\label{thm1} In the model (\ref{pdmodel}), we suppose $y_{i0}=0$ for all $i\ge 1$ and the innovations $\{\varepsilon_{it}, i\ge 1, t\ge 0\}$ are i.i.d. random variables with $E[\varepsilon_{11}]=0$ and $E[\varepsilon_{11}^2]=1$. In addition, we assume there exist two positive functions of $T$, Q(T) and P(T), such that
$$A_i^T:=P(T)\left(\sum_{t=1}^T y_{i,t-1} \varepsilon_{it} \right) \dconv A_i , \qquad T\rightarrow \infty,$$
and
$$B_i^T:=Q(T)\left( \sum_{t=1}^T y_{i,t-1}^2 \right) \dconv B_i,\qquad T\rightarrow \infty,$$
where $A_i$'s and $B_i$'s are random variables. \\
(1) If, as $T\rightarrow \infty$, $E[(A_i^T)^{r}]\rightarrow E[A_i^{r}]$ for $r=1, 2$ and $E[B_i^T]\rightarrow E[B_i]$ with $0<E[A_i^{2}]<\infty$ and $0<E[B_i]<\infty$ for all $i\ge 1$. Then we have
\begin{eqnarray}\label{2.5}
\sqrt{N}\frac{P(T)}{Q(T)}(\hat{\rho}-\rho) \dconv N\left(0,\frac{Var(A_1)}{(E[B_1])^2}\right), \qquad N, T\rightarrow \infty.
\end{eqnarray}
(2) If the conditions in (1) are fulfilled, and in addition,  as $T\rightarrow \infty$, $E[(B_i^T)^2] \rightarrow E[B_i^2]<\infty$ and $E[|A_i^T|^{3}]\rightarrow E[|A_i|^{3}]<\infty$ for all $i\ge 1$. Then we have, as long as $T$ is large enough, $\sqrt{N}\frac{P(T)}{Q(T)}(\hat{\rho}-\rho)$ converges to a normal distribution with the rate at least $O(N^{-\frac{1}{3}})$ as $N\rightarrow \infty$.
}
\\

\begin{remark}
We assume the cross section dimension $N$ and the time series dimension $T$ are independent in this paper. However, if $N$ depends on $T$ and is a monotonic function of $T$, one could still have the all results in this paper by some limit theorems for triangular arrays (for example, central limit theorem for triangular arrays in Levin and Lin (1992) and the law of large numbers for triangular arrays in Sung (1999)).
\end{remark}

\begin{proof}
(1) Apparently, $\{A_i^T, i\ge 1\}$ are i.i.d. random variables with $E[A_i^T]=0$. Moreover, it follows from the conditions of moment convergence that there exists some $T_0>0$ such that when $T>T_0$, $0<E[(A_i^T)^2]<\infty$. Denote
\begin{eqnarray}\label{2.7}
S_N^T=\frac{1}{\sqrt{N}}\sum_{i=1}^N \frac{A_i^T}{\sqrt{Var(A_1^T)}}.
\end{eqnarray}
Note that $E\left[\frac{A_i^T}{\sqrt{Var(A_1^T)}}\right]=0$ and $Var\left(\frac{A_i^T}{\sqrt{Var(A_1^T)}}\right)=1$. Hence, when $T>T_0$, applying the central limit theorem for i.i.d. random variables with zero mean and finite second moment leads to
\begin{eqnarray}\label{2.8}
S_N^T \dconv N(0,1), \qquad N\rightarrow \infty,
\end{eqnarray}
which, in view of characteristic function arguments,  further implies that
\begin{eqnarray}\label{2.28}
S_N^T \dconv N(0,1), \qquad N, T\rightarrow \infty.
\end{eqnarray}
In addition, noting that $\{B_i^T, i\ge 1\}$ are also i.i.d. random variables and there exists some $T_1>0$ such that $E[B_i^T]<\infty$ when $T>T_1$ by the conditions of moment convergence, it follows from the law of large numbers that when $T>T_1$,
\begin{eqnarray}\label{2.9}
\frac{1}{N}\sum_{i=1}^N B_i^T\pconv E[B_1^T], \qquad N\rightarrow \infty.
\end{eqnarray}
This easily yields
\begin{eqnarray}\label{2.29}
\frac{1}{N}\sum_{i=1}^N B_i^T\pconv E[B_1], \qquad N, T\rightarrow \infty.
\end{eqnarray}
Combining (\ref{2.28}) with (\ref{2.29}) immediately leads to (\ref{2.5}) by observing the following equality
\begin{eqnarray*}
\sqrt{N}\frac{P(T)}{Q(T)}(\hat{\rho}-\rho)=\sqrt{N}\frac{\sum_{i=1}^N A_i^T}{\sum_{i=1}^N B_i^T}=S_N^T\frac{\sqrt{Var(A_1^T)}}{\frac{1}{N}\sum_{i=1}^N B_i^T}.
\end{eqnarray*}

\noindent(2) It follows from the conditions of moment convergence that there exists some $T_2>0$ such that, when $T>T_2$, $E[|A_i^T|]^3<\infty$ and (\ref{2.8}) is still true. Denote
$$\gamma_T=E[|A_i^T|^3], \quad \sigma_T^2=E[(A_i^T)^2].$$
Then, according to the well-known Berry-Esseen bound for i.i.d. random variables with finite third moment, the speed of convergence for (\ref{2.8}), when $T>T_2$, is characterized by the following inequality:
\begin{align}\label{2.11}
\sup_x |P(S_N^T\leq x) - \Phi(x)| \leq \frac{c_0 \gamma_T}{\sigma_T^3\sqrt{N}},
\end{align}
where $c_0$ is some positive constant and $\Phi(x)$ is the distribution function of a standard normal random variable. In addition, by virtue of the conditions of moment convergence again, there esixts some $T_3>0$ such that $E[B_i^T]>0$ and $E[(B_i^T)^2]<\infty$ when $T>T_3$. Denote
$$ R_N^T= \frac{\frac{1}{N} \sum_{i=1}^N B_i^T }{E[B_1^T]}.$$
Note that $R_N^T$ is a non-negative random variable and $E[R_N^T]=1$. By applying Chebyshev's inequality, we have for any $0<\delta<\frac{1}{2}$,
\begin{align}\label{2.12}
P(|R_N^T-1|\geq \delta) \leq \frac{Var(R_N^T)}{\delta^2}=\frac{1}{N\delta^2} \frac{Var(B_1^T)}{(E[B_1^T])^2}.
\end{align}
Next, we will explore the convergence rate of $S_N^T/R_N^T$ for $T>\max\{T_2, T_3\}$.

First, when $x\geq 0$, one has
\begin{align}\label{2.13}
& \sup_{x\geq 0} \left| P\left( \frac{S_N^T}{R_N^T}<x \right) -\Phi(x) \right| \notag \\
= & \sup_{x\geq 0} \left| P\left( \frac{S_N^T}{R_N^T}<x , |R_N^T-1|<\delta \right)
    + P\left( \frac{S_N^T}{R_N^T}<x , |R_N^T-1|\geq \delta \right)-\Phi(x) \right| \notag \\
\leq &\sup_{x\geq 0} \left| P\left(S_N^T<R_N^T x, |R_N^T-1|<\delta  \right)
    -\Phi(x) \right| + P\left( |R_N^T-1|\geq \delta \right) \notag \\
\leq &\sup_{x\geq 0} \max{\left\{ P\left(S_N^T<(1+\delta) x \right)-\Phi(x), \Phi(x)-P\left(S_N^T<(1-\delta) x, |R_N^T-1|<\delta  \right)\right\}} \notag \\
    &+ P\left( |R_N^T-1|\geq \delta \right)\notag \\
\leq &\sup_{x\geq 0} \max{\left\{ P\left(S_N^T<(1+\delta) x \right)-\Phi(x), \Phi(x)-P\left(S_N^T<(1-\delta) x \right)+P(|R_N^T-1|\geq\delta)\right\}} \notag \\
    &+ P\left( |R_N^T-1|\geq \delta \right) \notag \\
\leq & \max{ \left\{ \sup_{x\geq 0}\left| P\left( S_N^T<(1+\delta)x \right)-\Phi((1+\delta)x)\right|
    +\sup_{x\geq 0}\left|\Phi((1+\delta)x)-\Phi(x) \right|+P\left( |R_N^T-1|\geq \delta \right), \right.}  \notag \\
    &\left. \sup_{x\geq 0} \left| P\left( S_N^T<(1-\delta)x \right)-\Phi((1-\delta)x)\right|
    +\sup_{x\geq 0}\left|\Phi((1-\delta)x)-\Phi(x) \right|+2P\left( |R_N^T-1|\geq \delta \right) \right\}.
\end{align}
Note that $\Phi(x)=\int_{-\infty}^{x} \frac{1}{\sqrt{2\pi}} e^{-\frac{t^2}{2}} dt$ satisfies the following smooth conditions:
\begin{align}\label{2.14}
\sup_{x\ge 0}|\Phi((1+\delta)x)-\Phi(x)| =\sup_{x\ge 0}\int_{x}^{(1+\delta)x}\frac{1}{\sqrt{2\pi}} e^{-\frac{t^2}{2}}dt \leq  \sup_{x\ge 0}\frac{\delta}{\sqrt{2\pi}}xe^{-\frac{x^2}{2}}  \leq \frac{\delta}{\sqrt{2\pi e}}
\end{align}
and similarly,
\begin{align}\label{2.16}
\sup_{x\ge 0}|\Phi(x)-\Phi((1-\delta)x)| \leq \frac{1}{1-\delta}\frac{\delta}{\sqrt{2\pi e}}<\frac{2\delta}{\sqrt{2\pi e}}.
\end{align}
Substituting (\ref{2.11}), (\ref{2.12}), (\ref{2.14}) and (\ref{2.16}) into (\ref{2.13}) and taking $\delta=N^{-\frac{1}{3}}$ ($N>8$), one has
\begin{eqnarray}\label{2.17}
\sup_{x\geq 0} \left| P\left( \frac{S_N^T}{R_N^T}<x \right) -\Phi(x) \right|
&\leq&   \frac{c_0 \gamma_T^3}{\sigma_T^3\sqrt{N}}+\frac{2}{N\delta^2} \frac{Var(B_1^T)}{(E[B_1^T])^2} +\frac{2\delta}{\sqrt{2\pi e}} \non
&=:&C_1(T)N^{-\frac{1}{2}}+ C_2(T) N^{-\frac{1}{3}},
\end{eqnarray}
where $C_1(T)=\frac{c_0 \gamma_T^3}{\sigma_T^3}$ and $C_2(T)=\frac{2 Var(B_1^T)}{(E[B_1^T])^2} +\frac{2}{\sqrt{2\pi e}}$. Note that both $C_1(T)$ and $C_2(T)$ are bounded when $T>\max\{T_2, T_3\}$.

By the same arguments, when $x<0$, one has
\begin{align}\label{2.18}
& \sup_{x<0} \left| P\left( \frac{S_N^T}{R_N^T}<x \right) -\Phi(x) \right| \notag \\
\leq & \max{ \left\{ \sup_{x<0}\left| P\left( S_N^T<(1-\delta)x \right)-\Phi((1-\delta)x)\right|
    +\sup_{x<0}\left|\Phi((1-\delta)x)-\Phi(x) \right|+P\left( |R_N^T-1|\geq \delta \right), \right.}  \notag \\
    &\left. \sup_{x<0} \left| P\left( S_N^T<(1+\delta)x \right)-\Phi((1+\delta)x)\right|
    +\sup_{x<0}\left|\Phi((1+\delta)x)-\Phi(x) \right|+2P\left( |R_N^T-1|\geq \delta \right) \right\} \notag  \\
\leq & C_1(T)N^{-\frac{1}{2}}+ C_2(T) N^{-\frac{1}{3}}.
\end{align}

Thus we can unify (\ref{2.17}) and (\ref{2.18}) as
\begin{align}\label{2.19}
\sup_{x\in R} \left| P\left( \frac{S_N^T}{R_N^T}<x \right) -\Phi(x) \right| \leq  C_1(T)N^{-\frac{1}{2}}+ C_2(T) N^{-\frac{1}{3}},
\end{align}
where both $C_1(T)$ and $C_2(T)$ are bounded when $T>\max\{T_2, T_3\}$.

Noting that
\begin{eqnarray}\label{2.10}
\sqrt{N}\frac{P(T)}{Q(T)}(\hat{\rho}-\rho)=\frac{S_N^T}{R_N^T}\cdot \frac{\sqrt{Var(A_1^T)}}{E[B_1^T]},
\end{eqnarray}
it is true that $\sqrt{N}\frac{P(T)}{Q(T)}(\hat{\rho}-\rho)$ also converges to a standard normal distribution with rate at least $O(N^{-\frac{1}{3}})$ as long as $T$ is large enough.
$\hfill \Box$
\end{proof}
\\

\begin{remark}
Generally the requirements of $0<E[A_i^2]<\infty$,$0<E[B_i]<\infty$ and convergence of moments are not strong They can be fulfilled in most of models we will discuss.
\end{remark}

Theorem 2.1 illustrates that $N$ determines the form of the limiting distribution while $T$ portrays the speed of convergence (with $P(T)$ and $Q(T)$). Considering the limiting distribution is normal, it can be totally depicted by its variance.
So the rest of this paper devotes to study the variance of the limiting distribution in various cases.

\section{Applications}
\setcounter{equation}{0}

In this section, the limiting distribution of the LSE of $\rho$ in model (\ref{pdmodel}) will be introduced one by one whenever $\rho$ is a fixed constant or a constant depending on $T$.

The results in the following lemma are taken from Mann and Wald (1943) and Rao (1978), respectively.

\begin{lemma}\label{lem1}
In the model (\ref{ar1}), we suppose $y_{0}=0$ and the innovations $\{\varepsilon_{t}, t\ge 1\}$ are i.i.d. random variables with $E[\varepsilon_{1}]=0$ and $E[\varepsilon_{1}^2]=1$. Then,\\
(1)~when $|\rho|<1$, one has
\begin{align*}
\sqrt{\frac{1-\rho^2}{T}}\sum_{t=1}^T y_{t-1}\varepsilon_{t} &\dconv  N(0,1), \qquad T\rightarrow \infty\\
\frac{1-\rho^2}{T}\sum_{t=1}^T y_{t-1}^2 &\pconv 1, \qquad T\rightarrow \infty;
\end{align*}
(2)~when $\rho=1$, one has
\begin{align*}
\left(\frac{1}{T}\sum_{t=1}^T y_{t-1}\varepsilon_{t}, \frac{1}{T^2}\sum_{t=1}^T y_{t-1}^2\right) &\dconv \left(\frac{1}{2}(W(1)^2-1), \int_0^1 W^2(t)dt\right),\qquad T\rightarrow \infty,
\end{align*}
where $\{W(t):0\leq t\leq 1 \}$ is a standard Wiener process.
\end{lemma}

Note that the LSE of $\rho$ in model (\ref{pdmodel}) is (\ref{pdlse}). With Theorem \ref{thm1} and Lemma \ref{lem1}, the following results can be gotten.
{\theorem\label{thm3}
In the model (\ref{pdmodel}), we suppose $y_{i0}=0$ for all $i\ge 1$ and the innovations $\{\varepsilon_{it}, i\ge 1, t\ge 0\}$ are i.i.d. random variables with $E[\varepsilon_{11}]=0$ and $E[\varepsilon_{11}^2]=1$. Then,\\
(1)~when $|\rho|<1$, one has
\begin{align*}
\frac{\sqrt{NT}}{\sqrt{1-\rho^2}}(\hat{\rho}-\rho) \dconv N(0,1), \qquad N,T \rightarrow \infty;
\end{align*}
(2)~when $\rho=1$, one has
\begin{align*}
\sqrt{N}T(\hat{\rho}-\rho) \dconv  N(0,2), \qquad N,T \rightarrow \infty.
\end{align*}
}

\proof (1) is easy and omitted. (2) is true because for any $i\ge 1$,
$$E\left[\frac{1}{T}\sum_{t=1}^T y_{i, t-1}\varepsilon_{t}\right]=0,\quad E\left[\left(\frac{1}{T}\sum_{t=1}^T y_{i, t-1}\varepsilon_{t}\right)^2\right]=\frac{1}{T^2}\sum_{t=1}^T(t-1)\rightarrow \frac{1}{2},$$
$$E\left[\frac{1}{T^2}\sum_{t=1}^T y_{t-1}^2\right]=\frac{1}{T^2}\sum_{t=1}^T (t-1)\rightarrow \frac{1}{2},$$
$$E\left[\frac{1}{2}(W(1)^2-1)\right]=0,\quad Var\Big(\frac{1}{2}(W(1)^2-1)\Big)=E\left[\left(\frac{1}{2}(W(1)^2-1)\right)^2\right]=\frac{1}{4}\times (3-1)=\frac{1}{2}$$
and
$$E\Big[\int_0^1 W^2(t)dt\Big]=\int_0^1 tdt=\frac{1}{2}.$$
$\hfill \Box$\\

\begin{remark}
The result (2) in Theorem \ref{thm3} is indeed one of the main results in Levin and Lin (1992), but the moment conditions in this paper are weaker than those in Levin and Lin (1992).
\end{remark}

\begin{remark}
In lemma \ref{lem1}, the case of $|\rho|>1$ is excluded. Anderson (1959) proved that, if $\{\varepsilon_{t}, t\ge 1\}$ are i.i.d. normal random variables with mean zeros and variance ones, then
\begin{align*}
\rho^{-(T-2)}\sum_{t=1}^T y_{t-1}\varepsilon_{t} &\dconv \xi\eta, \qquad T\rightarrow \infty,\\
(\rho^2-1)\rho^{-2(T-1)}\sum_{t=1}^T y_{t-1}^2 &\dconv \xi^2,\qquad T\rightarrow \infty,\\
\frac{\rho^T}{\rho^2-1}(\hat{\rho}-\rho)&\dconv C,
\end{align*}
where, $\xi$ and $\eta$ are independent and obey $N(0, \rho^2/(\rho^2-1))$ and $C$ stands for a standard Cauchy variate. In general case, $\hat{\rho}-\rho$ may not has a limiting distribution. Consequently, the case of $|\rho|>1$ is also excluded in Theorem \ref{thm3}.
\end{remark}

Next, we will study the case of $|\rho|>1$ in panel data AR(1) model without the help of Theorem \ref{thm1}.

{\theorem\label{thm4}
In the model (\ref{pdmodel}) with $|\rho|>1$, we suppose $y_{i0}=0$ for all $i\ge 1$ and the innovations $\{\varepsilon_{it}, i\ge 1, t\ge 0\}$ are i.i.d. random variables with $E[\varepsilon_{11}]=0$ and $E[\varepsilon_{11}^2]=1$. Then
\begin{eqnarray}\label{3.20}
\sqrt{N}\rho^{T-2}(\hat{\rho}-\rho) \dconv N(0,1), \qquad N,T \rightarrow \infty.
\end{eqnarray}
}

\begin{proof}
Denote $\beta=1/\rho$, and
\begin{eqnarray*}
&&u_{iT}=\varepsilon_{i1}+\beta \varepsilon_{i2}+\cdots +\beta^{T-2}\varepsilon_{i, T-1},\quad i\ge 1\\
&&v_{iT}=\varepsilon_{iT}+\beta \varepsilon_{i, T-1}\cdots +\beta^{T-2}\varepsilon_{i2}+\beta^{T-1}\varepsilon_{i1}, \quad i\ge 1.
\end{eqnarray*}
Then, following the proofs of Theorem 2.1 and Theorem 2.2 in Anderson (1959), one immediately has
\begin{eqnarray*}
&&\frac{1}{N}\left|\beta^{T-2}\sum_{i=1}^N \sum_{t=1}^T y_{i,t-1}\varepsilon_{it}-\sum_{i=1}^N u_{iT}v_{iT}\right|\pconv 0,\\
&&\frac{1}{N}\left|\beta^{2(T-2)}\sum_{i=1}^N\sum_{t=1}^T y_{i,t-1}^2-\sum_{i=1}^N u_{iT}^2\right|\pconv 0.
\end{eqnarray*}
As a result,
\begin{eqnarray}\label{3.15}
\sqrt{N}\rho^{T-2}(\hat{\rho}-\rho)&=&\sqrt{N}\rho^{T-2}\frac{\sum_{i=1}^N \sum_{t=1}^T y_{i,t-1}\varepsilon_{it}}{\sum_{i=1}^N\sum_{t=1}^T y_{i,t-1}^2}\non
&=&\sqrt{N}\frac{\beta^{T-2}\frac{1}{N}\sum_{i=1}^N \sum_{t=1}^T y_{i,t-1}\varepsilon_{it}}{\beta^{2(T-2)}\frac{1}{N}\sum_{i=1}^N\sum_{t=1}^T y_{i,t-1}^2}\non
&=&\frac{\frac{1}{\sqrt{N}}\sum_{i=1}^N u_{iT}v_{iT}}{\frac{1}{N}\sum_{i=1}^N u_{iT}^2}(1+o_P(1)).
\end{eqnarray}
That is to say, we only need to derive the limiting distribution of $\frac{\frac{1}{\sqrt{N}}\sum_{i=1}^N u_{iT}v_{iT}}{\frac{1}{N}\sum_{i=1}^N u_{iT}^2}$ in order to derive the limiting distribution of $\sqrt{N}\rho^{T-2}(\hat{\rho}-\rho)$.

First, for any fixed $T\ge 2$, it follows from the law of large numbers that
\begin{eqnarray}\label{3.16}
\frac{1}{N}\sum_{i=1}^N u_{iT}^2\pconv \frac{1-\beta^{2(T-1)}}{1-\beta^2},\quad N\rightarrow \infty,
\end{eqnarray}
which yields
\begin{eqnarray}\label{3.21}
\frac{1}{N}\sum_{i=1}^N u_{iT}^2\pconv \frac{1}{1-\beta^2},\quad N, T\rightarrow \infty.
\end{eqnarray}
Second, denote
\begin{eqnarray*}
u_{iT}^{*}=\sum_{t=1}^{[T/2]}\beta^{t-1}\varepsilon_{it},\quad v_{iT}^{*}=\sum_{t=[T/2]+1}^{T}\beta^{T-t}\varepsilon_{it},
\end{eqnarray*}
here the symbol $[x]$ denote the largest integer not greater than $x$. Then, by the proof of Theorem 2.3 in Anderson (1959), we have for any $i\ge 1$,
$$|u_{iT}-u_{iT}^{*}|\pconv 0~\text{and}~|u_{iT}-u_{iT}^{*}|\pconv 0, \quad T\rightarrow \infty.$$
It follows that
\begin{eqnarray}\label{3.17}
\frac{1}{\sqrt{N}}\sum_{i=1}^N u_{iT}v_{iT}=\frac{1}{\sqrt{N}}\sum_{i=1}^N u_{iT}^{*}v_{iT}^{*}(1+o_P(1)).
\end{eqnarray}
Note that the sequences $\{u_{iT}^{*}, i\ge 1\}$ and $\{v_{iT}^{*}, i\ge 1\}$ are independent for any fixed $T\ge 2$. Then, by virtue of the central limit theorem we have
\begin{eqnarray}\label{3.23}
\frac{1}{\sqrt{N}}\sum_{i=1}^N u_{iT}^{*}v_{iT}^{*}\dconv N(0, \frac{(1-\beta^{2[T/2]})(1-\beta^{2(T-[T/2])})}{(1-\beta^2)^2}), \quad N\rightarrow \infty,
\end{eqnarray}
which further implies that
\begin{eqnarray}\label{3.18}
\frac{1}{\sqrt{N}}\sum_{i=1}^N u_{iT}^{*}v_{iT}^{*}\dconv N(0, \frac{1}{(1-\beta^2)^2}), \quad N,T\rightarrow \infty
\end{eqnarray}
by characteristic function arguments. Now, combining (\ref{3.15}), (\ref{3.21}), (\ref{3.17}) with (\ref{3.18}) yields (\ref{3.20}). $\hfill \Box$
\end{proof}
\\

\begin{remark}
It is interesting to see that $\hat{\rho}-\rho$ has a limiting distribution in panel data case while it may not has a limiting distribution in univariate time series case.
\end{remark}

Though in penal data, the form of limiting distribution is stable, noticing that the scale of
$\hat{\rho}-\rho$ declines from $O(\frac{1}{\sqrt{NT}})$ when $|\rho|<1$, to $O(\frac{1}{\sqrt{N}T})$ when $\rho=1$ and to $O(\frac{1}{\sqrt{N}\rho^{T-2}})$ when $|\rho|>1$, the rate of convergence changes radically at $\rho=1$. Hence, it is necessary to discuss the case when $\rho$ is near $1$. In the rest of the paper, we suppose $\rho$ depends on $T$, so it is natural to use the notation $\hat{\rho}_T$ to denote the LSE of $\rho$, that is, (\ref{pdlse}).

We first follow the proposal of Chan and Wei (1987) and Phillips (1987) to study the case where $\rho=\rho_T=1-\frac{c}{T}$, where $c$ is a fixed constant.
Consider the model
\begin{align*}
y_{it}=\rho_T y_{i,t-1}+\varepsilon_{it},\qquad \rho_T=1-\frac{c}{T},\qquad t=1,2,...,T, \qquad i=1,2,...,N,
\end{align*}
where $y_{i0}=0$ for all $i\ge 1$ and $\{\varepsilon_{it}, i\ge 1, t\ge 1\}$ are i.i.d. random variables with $E[\varepsilon_{11}]=0$ and $E[\varepsilon_{11}^2]=1$.

The following lemma is not explicitly formulated in Chan and Wei (1987), but can be easily obtained by the proofs in Chan and Wei (1987), technique of change of variable and It\'{o} formula. Thus, the details are omitted here.

\begin{lemma}\label{lem2}
Let $\rho_T=1-\frac{c}{T}$, where $c\neq 0$ is a fixed constant. Suppose $y_{t}$ comes from the following reparameterized AR(1) model,
$$y_{t}=\rho_T y_{t-1}+\varepsilon_{t}, \qquad t=1,2,...T,$$
where $y_0=0$ and $\{\varepsilon_{t}, t\ge 1\}$ are i.i.d. random variables with $E[\varepsilon_{1}]=0$ and $E[\varepsilon_{1}^2]=1$. Then
\begin{eqnarray}
&&T^{-1}\sum_{t=1}^T y_{t-1} \varepsilon_t \dconv \frac{b}{2c}\int_0^1 (1+bt)^{-1}W(t)dW(t),\label{3.1}\\
&&T^{-2}\sum_{t=1}^T y_{t-1}^2 \dconv \left(\frac{b}{2c}\right)^2\int_0^1 (1+bt)^{-2}W^2(t)dt,\label{3.2}
\end{eqnarray}
where $b=e^{2c}-1$ and $\{W(t), 0\leq t\leq 1\}$ is a standard Wiener process.
\end{lemma}

With the help of the above lemma and Theorem \ref{thm1}, we have the following result.

{\theorem\label{thm2}
Let $\rho=\rho_T=1-\frac{c}{T}$, where $c\neq 0$ is a fixed constant. For $t=1,2,...,T$ and $i=1,2,...,N$, we suppose $y_{it}$ satisfied the following reparameterized AR(1) model,
$$y_{it}=\rho_T y_{i,t-1}+\varepsilon_{it}, \qquad i=1,2,...,N,\quad t=1,2,...T,$$
where $y_{i0}=0$ for all $i\ge 1$ and $\{\varepsilon_{it}, i\ge 1, t\ge 1\}$ are i.i.d random variables with $E[\varepsilon_{11}]=0$ and $E[\varepsilon_{11}^2]=1$. Then
\begin{align}\label{3.11}
\sqrt{N}T(\hat{\rho}_T-\rho_T)\dconv N\left(0, \frac{4c^2}{2c-1+e^{-2c}}\right).
\end{align}
}

\begin{proof}
It follows from Theorem \ref{thm1} and Lemma \ref{lem2} that we only need to verify the corresponding conditions of moment convergence and calculate the variance of the right hand side of (\ref{3.1}) and the expectation of the right hand side of (\ref{3.2}). To verify the conditions of moment convergence. It is true that for every $i\ge 1$
$$E\left[T^{-1}\sum_{t=1}^T y_{i, t-1} \varepsilon_{it}\right]=0,$$
\begin{eqnarray*}
 E\left[\left(T^{-1}\sum_{t=1}^T y_{i, t-1} \varepsilon_{it}\right)^2\right]&=&\frac{1}{T^2}\sum_{t=1}^T\frac{1-\rho_T^{2(t-1)}}{1-\rho_T^2}\\
 &=&\frac{1}{T^2(1-\rho_T^2)}\left(T-\frac{1-\rho_T^{2T}}{1-\rho_T^2}\right),\\
 &\rightarrow&\frac{2c-1+e^{-2c}}{4c^2},
\end{eqnarray*}
$$E\left[T^{-2}\sum_{t=1}^T y_{i, t-1}^2\right]=E\left[\left(T^{-1}\sum_{t=1}^T y_{i, t-1} \varepsilon_{it}\right)^2\right]\rightarrow\frac{2c-1+e^{-2c}}{4c^2},$$
$$E\left[\frac{b}{2c}\int_0^1 (1+bt)^{-1}W(t)dW(t)\right]=0,$$
\begin{eqnarray*}
E\left[\left(\frac{b}{2c}\int_0^1 (1+bt)^{-1}W(t)dW(t)\right)^2\right]&=& \frac{b^2}{4c^2}E\left[\int_0^1 \left((1+bt)^{-1}W(t) \right)^2dt\right] \\
&=&\frac{b^2}{4c^2}\int_0^1 \frac{t}{(1+bt)^2}dt\\
&=&\frac{1}{4c^2}\left[\ln(1+b)-b/(1+b)\right]\\
&=&\frac{2c-1+e^{-2c}}{4c^2}
\end{eqnarray*}
according to It$\acute{o}$ isometry theorem, and
\begin{eqnarray*}
E\left[\left(\frac{b}{2c}\right)^2\int_0^1 (1+bt)^{-2}W^2(t)dt\right]&=&E\left[\left(\frac{b}{2c}\int_0^1 (1+bt)^{-1}W(t)dW(t)\right)^2\right]\\
&=&\frac{2c-1+e^{-2c}}{4c^2}.
\end{eqnarray*}
To calculate the variance of the right hand side of (\ref{3.1}) and the expectation of the right hand side of (\ref{3.2}). Note that the latter has just been done. For the former, it is easy to see that
\begin{eqnarray*}
Var\left(\frac{b}{2c}\int_0^1 (1+bt)^{-1}W(t)dW(t)\right)&=& E\left[\left(\frac{b}{2c}\int_0^1 (1+bt)^{-1}W(t)dW(t)\right)^2\right] \\
&=& \frac{2c-1+e^{-2c}}{4c^2}.
\end{eqnarray*}
The proof is complete. $\hfill \Box$\\
\end{proof}

\begin{remark}
It is easy to see that
$$\lim_{c\rightarrow 0}\frac{4c^2}{2c-1+e^{-2c}}=2.$$
Thus, the second part of Theorem \ref{thm3} can be regarded as a complementary of Theorem \ref{thm2}.
\end{remark}

\bigskip

Now we investigate the case where $\rho=1-\frac{c}{k_T}$, where $c\neq 0$ and $k_T=o(T)$ is an increasing function of $T$ diverging to infinity. First, we introduce a result about the limiting distribution of $\hat{\rho}_T$ in univariate time seires AR(1) model which is taken from Philips and Magdalinos (2007).

\begin{lemma}\label{lem3}
Let $\rho_T=1-\frac{c}{k_T}$, where $c\neq 0$ is a fixed constant and $k_T=o(T)$ is an increasing function of $T$ diverging to infinity. For $t=1,2,...,T$, suppose $y_t$ satisfied the following reparameterized AR(1) model,
$$y_t=\rho_T y_{t-1}+\varepsilon_t, \qquad t=1,2,...T,$$
where $y_0=0$ and $\{\varepsilon_t, t\ge 1\}$ are i.i.d. random variables with $E[\varepsilon_{1}]=0$ and $E[\varepsilon_{1}^2]=1$. Then, for $c>0$,
\begin{eqnarray}\label{3.9}
\frac{1}{\sqrt{Tk_T}}\sum_{t=1}^T y_{t-1} \varepsilon_t \dconv N(0, \frac{1}{2c}),
\end{eqnarray}
\begin{eqnarray}\label{3.10}
\frac{1}{Tk_T}\sum_{t=1}^T y_{t-1}^2 \pconv  \frac{1}{2c};
\end{eqnarray}
and for $c<0$,
\begin{eqnarray}\label{3.14}
\left(\frac{1}{\rho_T^T k_T}\sum_{t=1}^T y_{t-1}\varepsilon_t, \frac{-2c}{(\rho_T^T k_T)^2} \sum_{t=1}^T y_{t-1}^2 \right) \dconv
    (XY,Y^2),
\end{eqnarray}
where $X$ and $Y$ are independent random variables obeying $N(0, 1/(-2c))$.
\end{lemma}

We now study the panel data case.
By employing Theorem \ref{thm1} and Lemma \ref{lem3}, we immediately have the following result.

{\theorem
Let $\rho_T=1-\frac{c}{k_T}$, where $c\neq 0$ is a fixed constant and $k_T=o(T)$ is an increasing function of $T$ diverging to infinity. For $t=1,2,...,T$ and $i=1,2,..,N$, suppose $y_{it}$ satisfied the following reparameterized AR(1) model,
$$y_{it}=\rho_T y_{i,t-1}+\varepsilon_{it}, \qquad t=1,2,...,T,\quad i=1,2,...,N,$$
where $y_{i0}=0$ for all $i\ge 1$ and $\{\varepsilon_{it}, i\ge 1, t\ge 1\}$ are i.i.d. random variables with $E[\varepsilon_{11}]=0$ and $E[\varepsilon_{11}^2]=1$. Then, for $c>0$ we have
\begin{eqnarray}\label{3.12}
\sqrt{NTk_T}(\hat{\rho_T}-\rho_T) \dconv N(0, 2c)
\end{eqnarray}
and for $c<0$ we have
\begin{eqnarray}\label{3.13}
\sqrt{N}k_T\rho_T^T(\hat{\rho}-\rho)\dconv N(0, 4c^2).
\end{eqnarray}
}
\\

\begin{proof}
The proofs of (\ref{3.12}) and (\ref{3.13}) are similar, so we only prove (\ref{3.13}) here. To do so, it follows from Theorem \ref{thm1} and Lemma \ref{lem2} that we only need to verify the corresponding conditions of moment convergence and calculate the variance of $XY$ and the expectation of $Y^2$ in the right hand side of (\ref{3.14}). Noting that $\rho_T^{-T}=o(k_T/T)$ by Proposition A.1 in Phillips and Magdalinos (2007), it is true that for every $i\ge 1$,
$$E\left[\frac{1}{\rho_T^T k_T}\sum_{t=1}^T y_{i, t-1}\varepsilon_{it}\right]=0,$$
\begin{eqnarray*}
 E\left[\left(\frac{1}{\rho_T^T k_T}\sum_{t=1}^T y_{i, t-1}\varepsilon_{it}\right)^2\right]&=&\frac{1}{\rho_T^{2T} k_T^2}\sum_{t=1}^T\frac{1-\rho_T^{2(t-1)}}{1-\rho_T^2}\\
 &=&\frac{1}{\rho_T^{2T} k_T^2(1-\rho_T^2)}\left(T-\frac{1-\rho_T^{2T}}{1-\rho_T^2}\right)\\
 &=&o(1)+\frac{1}{k_T^2(1-\rho_T^2)^2}\\
 &\rightarrow&\frac{1}{4c^2},
\end{eqnarray*}
$$E\left[\frac{-2c}{(\rho_T^T k_T)^2} \sum_{t=1}^T y_{i, t-1}^2\right]=\frac{-2c}{\rho_T^{2T} k_T^2}\sum_{t=1}^T\frac{1-\rho_T^{2(t-1)}}{1-\rho_T^2}=o(1)+\frac{-2c}{k_T^2(1-\rho_T^2)^2}\rightarrow -\frac{1}{2c},$$
$$E\left[XY\right]=0,\quad Var\left((XY)^2\right)=E\left[(XY)^2\right]=\frac{1}{4c^2},\quad E[Y^2]=-\frac{1}{2c}.$$
The proof is complete. $\hfill \Box$\\
\end{proof}
\\


\begin{thebibliography}{99}
\footnotesize

\bibitem{r1}
\textsc{Anderson.T.W.} (1959).
On asymptotic distributions of estimates of parameters of stochastic difference equations.
\textit{Ann. Math. Statist.},
\textbf{1}, 676-687.

\bibitem{r2}
\textsc{Chan, N. H. and Wei, C. Z.} (1987).
Asymptotic inference for nearly nonstationary AR (1) processes.
\textit{Ann. Statist.},
\textbf{1}, 1050-1063.

\bibitem{r3}
\textsc{Levin, A. T. and Lin, C. F.} (1992).
Unit root tests in panel data: asymptotic and finite-sample properties.
\textit{Economics Working Paper Series}.

\bibitem{r4}
\textsc{Mann, H. B. and Wald, A. } (1943).
On the statistical treatment of linear stochastic difference equations.
\textit{Econometrica},
\textbf{11}, 173-220.



\bibitem{r9}
\textsc{Phillips, P. C. B.} (1987).
Towards a unified asymptotic theory for autoregression.
\textit{Biometrika},
\textbf{74} (3), 535-547.


\bibitem{r6}
\textsc{Phillips, P. C. B. and Magdalinos, T.} (2007).
Limit theory for moderate deviations from a unit root.
\textit{Journal of Econometrics},
\textbf{1}, 115-130.



\bibitem{r7}
\textsc{Rao, M. M. } (1978).
Asymptotic distribution of an estimator of boundary parameter of an unstable process.
\textit{Ann. Statist.},
\textbf{6}, 185-190.

\bibitem{}
\textsc{Sung, S. H. } (1999).
Weak law of large numbers for arrays of random variables.
\textit{Statistics $\&$ Probability Letters},
\textbf{42}, 293-298.


\bibitem{r8}
\textsc{White, J. S. } (1958).
The limiting distribution of the serial correlation coefficient in the explosive case.
\textit{Ann. Math. Statist.},
\textbf{29}, 1188-1197.

\end{thebibliography}
\end{document}